\documentclass[11pt]{article}
\usepackage[margin=1in]{geometry}
\usepackage{amsmath,amssymb,amsfonts}
\usepackage{graphicx}
\usepackage{float}
\usepackage{caption}
\usepackage{subcaption}
\usepackage{multirow}
\usepackage{booktabs}
\usepackage{authblk}
\usepackage{array}
\usepackage{hyperref}
\usepackage{cleveref}
\usepackage{cite}
\usepackage{color}
\usepackage{algorithm}
\usepackage{algorithmic}
\usepackage{bm}
\usepackage[numbers,sort&compress]{natbib}
\usepackage{url}     
\usepackage{doi}     

\title{Acoustic Holography in the Megahertz Frequency Range with Optimal Lens Topologies and Nonlinear Acoustic Feedback}

\author[1*]{Pradosh~P.~Dash}
\author[1,2]{Costas~D.~Arvanitis}

\affil[1]{Mechanical Engineering, Georgia Institute of Technology,}
\affil[2]{Biomedical Engineering, Georgia Institute of Technology and Emory University}

\date{}

\begin{document}

\maketitle

\begin{abstract}
Acoustic holography in the megahertz frequency range can impact numerous applications, including manufacturing, non-destructive testing, and transcranial ultrasound. However, designing lens topologies for complex acoustic holograms in the megahertz range poses a significant challenge, as wave propagation effects through the lens cannot be ignored. Here, we show that the inherent ability of heterogeneous angular spectrum approach to incorporate in plane varying speed-of-sound maps and support rapid differentiable optimization of lens thickness profiles can generate lens topologies for high fidelity acoustic holography. Crucially, we show that this framework can also account for wavefront aberrations in the propagation media, providing the opportunity to reconfigure this disruptive technology for high precision neuro-interventions. Our investigations also revealed that low frequency acoustic feedback generated by nonlinear mixing of high frequency waves allows attaining accurate skull-compensating lens alignment and creates the possibility to monitor CSF fluid build-up and removal in hydrocephalus. Together, our findings support the design of simple, economical, and high-performance ultrasound systems.
\end{abstract}

\section{Introduction}

Recent advancements in acoustic holography based on holographic lenses offer a promising pathway for designing simpler, economical, and more flexible ultrasound systems for a range of applications, including contactless manufacturing~\cite{melde2023compact}, consumer electronics~\cite{hirayama2019volumetric}, non-destructive testing~\cite{xie2016acoustic}, imaging~\cite{kruizinga2017compressive}, and targeted neuro-interventions~\cite{jimenez2018adaptive,maimbourg2018printed,jimenez2019holograms}, among others~\cite{melde2016holograms,hu2022airy,andres2023holographic}. These acoustic holograms, also known as phase plates, encode spatial phase patterns onto passive lenses, effectively transforming a single-element transducer into one capable of generating complex volumetric pressure fields~\cite{jimenez2018adaptive,melde2016holograms}. This is possible because their apparent element (i.e., pixel) size is significantly smaller than one wavelength~\cite{melde2016holograms}. Crucially, for large apertures (i.e., several cm$^2$) this is equivalent to dense phased arrays (e.g., $10^4 - 10^5$ elements) that are unrealizable, creating the possibility to generate acoustic fields that based on current technologies are not considered to be possible. Moreover, recent implementations that employ time domain methods of wave propagation in heterogeneous media, such as the human skull, can account for skull-induced aberration to produce focal spots~\cite{maimbourg2018printed,daniel2024multifrequency} or pressure fields to concurrently target different brain regions~\cite{jimenez2019holograms,jimenez2024feasibility,he2022multitarget,kook2023multifocal,yao2025acoustic}.

While this approach can achieve the desired phase distribution, converting the optimized phase pattern to lens topology by scaling with the relative wavenumber, it implies that the hologram is a thin element that alters only the phase while neglecting amplitude changes and wave propagation effects within the lens. However, for high frequency systems ($\approx$ 1 MHz), which are critical for high fidelity holography (i.e., rich in information content), the lens feature size approaches the acoustic wavelength ($\lambda < 1.5$ mm) that invalidates the thin-element approximation, resulting in significant thickness-dependent amplitude and phase errors.

In addition to high frequencies, high-frequency holography requires extended optimization to prevent on- or off-target hot spots. Unfortunately, current frequency domain methods that are fast do not account for heterogeneities in the wave propagation~\cite{melde2016holograms} and time-domain simulations are computationally intensive (can take up several hours~\cite{jimenez2019holograms}). Also, for applications that require large apertures and the generation of holographic fields far from the transducer surface, such as transcranial ultrasound, are time-domain methods impractical and may hinder identifying optimal system configurations (i.e., require testing multiple configurations)~\cite{choi2024neuronavigation}. While hologram design can be accelerated using automatic differentiation methods~\cite{fushimi2021acoustic}, rapid 3D printed lens design free of thickness-dependent amplitude and phase errors (i.e., diffraction-limited holographic beam shaping) remains an important scientific and engineering challenge~\cite{jimenez2024feasibility}.

Our ability to reconfigure this potentially disruptive technology for biomedical applications, such as transcranial ultrasound, also hinges on our ability to accurately register the holographic lens to the patient's anatomy, orientation, and position relative to the transducer/lens plane~\cite{maimbourg2018printed,jimenez2024feasibility}. For instance, complex hologram designs and pressure field topologies, which require higher operation frequency ($\geq 0.7$ MHz) and skull-compensating lens topologies, are very sensitive to lens-skull misalignment~\cite{andres2022numerical}. Hence, high quality registration (i.e., sub-wavelength accuracy) is needed to preserve their fidelity and targeting accuracy. Unfortunately, at these frequencies non-MRI based registration methods cannot achieve the required alignment (sub-wavelength) that can lead to targeting errors of a few millimeters (i.e., 1 - 2 wavelengths) and reduced performance~\cite{choi2024neuronavigation,chen2021neuronavigation,wei2013neuronavigation,pouliopoulos2020clinical}. Therefore, robust and accurate registration strategies that can accurately align the lens to the patient's skull anatomy are critical for designing cost-effective and portable transcranial ultrasound (TUS) systems for high precision (i.e., sub-wavelength) neuro-interventions.

In this study, we introduce a framework for high fidelity acoustic holography. This framework takes advantage of the heterogeneous angular spectrum approach (HASA) -- a fast spectral method for wave propagation in complex media~\cite{schoen2020heterogeneous} -- inherent ability to incorporate in plane varying speed-of-sound maps and support differentiable optimization based on ADAM iterative optimizer to design acoustic holograms with complex topology. By employing mathematical modeling and experimental studies we demonstrate the abilities of the HASA-ADAM framework to design holographic lenses that encode complex acoustic holograms as well as aberration-free focusing through human skull. Subsequently, we leverage the parametric array (PA) signal -- a distinct low frequency acoustic signal generated by the nonlinear mixing of high frequency waves~\cite{westervelt1963parametric} -- to attain accurate skull-compensating lens alignment. Our investigations reveal that misaligned lens (i.e., aberrated) augments the finite-amplitude wave propagation effects within the highly non-linear skull, giving rise to strong PA signal that can penetrate the skull with minimal losses. Thus, minimizing the PA signal leads to an effective acoustic feedback mechanism to noninvasively align the holographic lens to the skull. Finally, we show that the PA signal can also be used to monitor changes in the effective nonlinearity of the brain in response to relative changes in ventricular size during hydrocephalus progression or shunt treatment~\cite{hochstetler2022hydrocephalus}.

\section{Results}

\subsection{HASA combined with ADAM iterative optimizer can design holographic lens topologies for high fidelity acoustic holography}

\begin{figure}[htbp]
\centering
\includegraphics[width=0.75\textwidth]{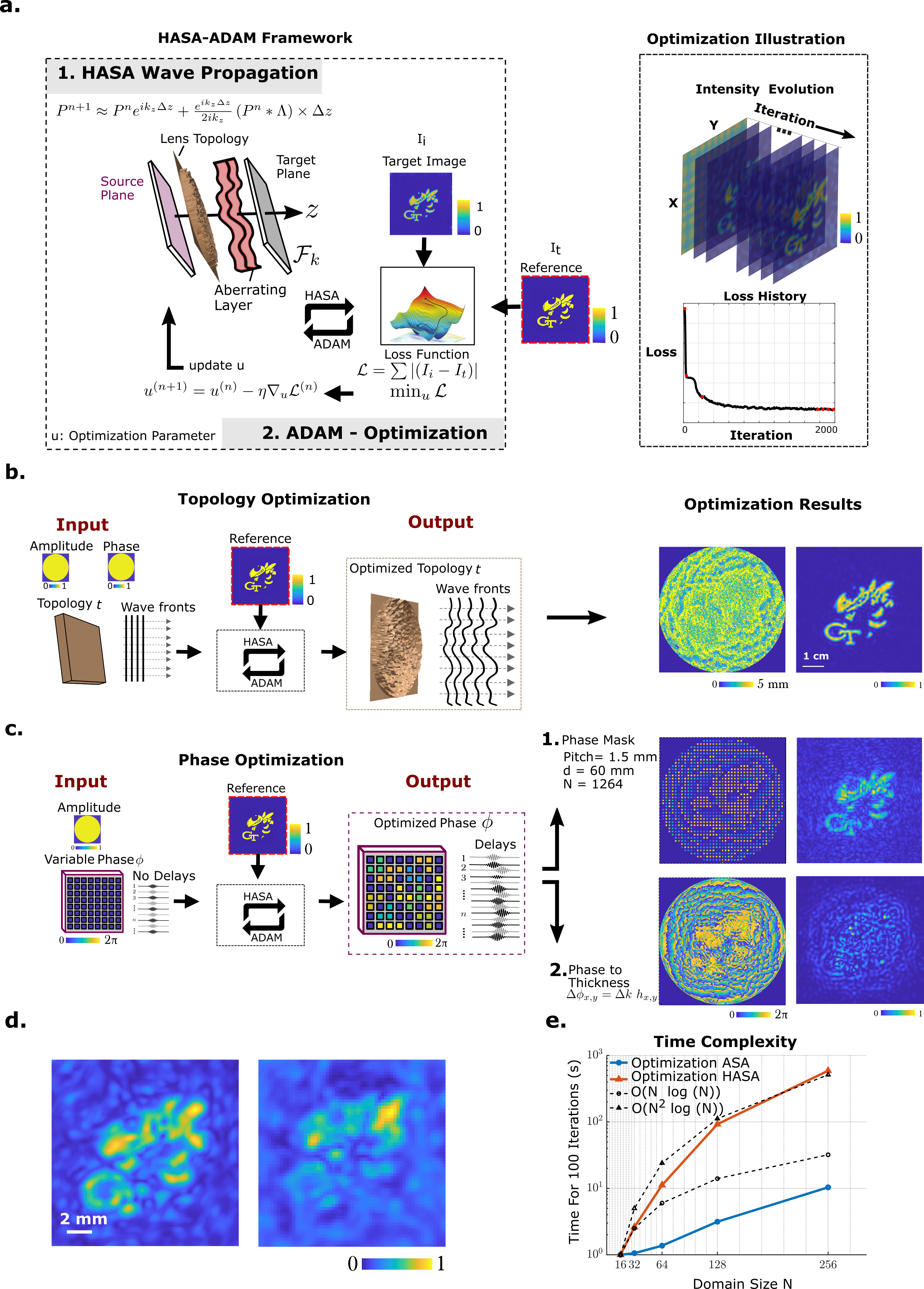}
\caption{HASA-ADAM provides a unified framework to design 3D high-fidelity, large aperture, acoustic holography. (a) Schematic of the hologram design framework, comprising two key steps: 1) Heterogeneous Angular Spectrum Approach (HASA) for wave propagation and 2) optimization using ADAM. The HASA algorithm for phase or thickness optimization via gradient descent is shown in Table~\ref{tab:algorithm}. (b) HASA-ADAM framework is used to optimize the topology of a lens for complex 2D hologram in free field (left), this results in PSNR 18.58 dB (right). (c) HASA-ADAM optimization results for complex 2D hologram illustrates the free field hologram optimized for a phased array with 1264 elements and $\lambda$ spacing with PSNR 17.5dB (top) gets severely deteriorated to 11.8 dB when a phase optimization lens is used (middle). (d) k-wave simulation results for topology optimized lens (left) and hydrophone scan for validation (right). (e) Complexity analysis of hologram optimization using HASA.}
\label{fig:hasa_adam_framework}
\end{figure}

We introduce a framework that combines the Heterogeneous Angular Spectrum Approach (HASA) -- a fast spectral method for wave propagation in complex media~\cite{schoen2020heterogeneous} -- with the ADAM iterative optimizer to reduce a loss function (absolute difference in intensity between reference or target image and image plane)~\cite{fushimi2021acoustic}, for accelerated hologram optimization (Fig.~\ref{fig:hasa_adam_framework}a and Table~\cref{tab:algorithm}). This approach takes advantage of HASA's inherent ability to incorporate in plane varying speed-of-sound maps and support a differentiable optimization of lens thickness profiles (i.e., assuming an initial zero phase and constant amplitude at the source; Fig.~\ref{fig:hasa_adam_framework}a and Table~\ref{tab:algorithm}). As a result, the proposed framework allows to design holographic lens topologies that account for the physical effects of wave propagation within the lens and, consequently, generate holographic lenses that encode complex acoustic holograms at the megahertz frequency range (Fig.~\ref{fig:hasa_adam_framework}b). This approach can also be used for acoustic holography based on optimized phase (Fig.~\ref{fig:hasa_adam_framework}c). 

\begin{table}[htbp]

\begin{center}
\caption{Table 1: HASA-ADAM algorithm for a) topology optimization, b) phase optimization}
\label{tab:algorithm}
\begin{tabular}{|l|l|}
\hline
a. HASA-ADAM: Topology Optimization & b. HASA-ADAM: Phase Optimization \\
\hline
\( \begin{aligned} & \text { Input: } N_{\max }, \epsilon, \eta, \lambda, u^{(0)}, I_{t} \\ & \text { Output: } t, \mathcal{L} \\ & \text { while } n<N_{\max } \text { and } \mathcal{L}^{(n)}>\epsilon \text { do } \\ & t^{(n)}=\min \left\{\operatorname{softplus}\left(u^{(n)}\right), t_{\max }\right\} \\ & P^{(n)}=\mathcal{H}\left(P_{0}, t^{(n)}\right) \\ & I^{(n)}=\left|P^{(n)}\right|^{2} \\ & \mathcal{L}^{(n)}=\sum_{x, y}\left|I_{i}^{(n)}(x, y)-I_{t}(x, y)\right|+\lambda R\left(t^{(n)}\right) \\ & u^{(n+1)}=u^{(n)}-\eta \nabla_{u} \mathcal{L}^{(n)} \\ & n \leftarrow n+1 \\ & \text { end return } t^{(n)}, \mathcal{L}^{(n)} \end{aligned} \) & \( \begin{aligned} & \text { Input: } N_{\max }, \epsilon, \eta, \lambda, P_{0}, P_{t} \\ & \text { Output: } \phi, \mathcal{L} \\ & \text { while } n<N_{\max } \text { and } \mathcal{L}^{(n)}>\epsilon \text { do } \\ & P_{i}^{(n)} \leftarrow \mathcal{H}\left(P_{0}, \phi^{(n)}\right) ; I_{i}^{(n)}=\left|P_{i}^{(n)}\right|^{2} \\ & \mathcal{L}^{(n)} \leftarrow \sum_{x, y}\left|I_{i}^{(n)}-I_{t}\right|+\lambda \mathcal{R}\left(\phi^{(n)}\right) ; \\ & \phi^{(n+1)} \leftarrow \phi^{(n)}-\eta \nabla_{\phi} \mathcal{L}^{(n)} ; \\ & P_{0}^{(n+1)} \leftarrow \alpha_{h} P_{0}^{(n)} ; n \leftarrow n+1 ; \\ & \text { end } \\ & \text { return } \phi^{(n)}, \mathcal{L}^{(n)} \end{aligned} \) \\
\hline
\end{tabular}
\end{center}

\end{table}

To test this concept and identify optimal method for high-fidelity acoustic holography, we aimed at generating a complex holographic pattern based on thickness optimization and compared it with holographic methods that rely on converting the optimized phase pattern to a thickness pattern (i.e. thin-element approximation) as well as a realizable, large aperture, phased array (Fig.~\ref{fig:hasa_adam_framework}c). HASA-ADAM can achieve better pattern definition and performance as compared to a realizable phase array with the same aperture (17.5 vs 18.5 dB PSNR). Moreover, the lens design based on current holographic lens design paradigms (i.e., based on thin-element approximation) was not able to produce the pattern. Physically accurate simulations and experimental measurements verified this performance that has so far been fundamentally limited by thickness-dependent amplitude and phase errors (Fig.~\ref{fig:hasa_adam_framework}d).

While the computational complexity ($\sim O(N^2\log N)$) of the proposed framework is higher from the homogeneous ASA ($\sim O(N\log N)$ (Fig.~\ref{fig:hasa_adam_framework}e)~\cite{melde2016holograms}, the optimized hologram phase converges in approximately 20 minutes using a discretization of $\lambda/10$ at 1 MHz within a domain size of N = 40 mm (corresponding to a volume of 40 mm$^3$), on a system equipped with a 24 GB NVIDIA RTX 3090 GPU. Moreover, by increasing the linear dimension by 50\% (i.e., 6 cm, which was upper limit in the capacity of the GPU used) results in about threefold increase in computational time (close to the 2.4 fold expected from the $O(N^2\log N)$ scaling and consistent with GPU-memory overhead), highlighting the method's scalability and efficiency for large aperture (i.e., clinical-scale) designs. Note that a major reduction in optimization time can be achieved by down sampling the domain to a discretization of $\lambda/6$ at 1 MHz without any loss in performance ($\sim$15 minutes for 500 iterations). Thus, we adopted $\lambda/6$ discretization for the subsequent studies in the paper. Together, HASA-ADAM constitutes a major advancement in acoustic holography, as it provides a unified framework to design 3D-printed lenses for high-fidelity acoustic holography, thereby creating the possibility to design high performance systems at the fraction of the cost.

\subsection{HASA combined with ADAM iterative optimizer can design holographic lens topologies for high performance TUS}

\begin{figure}[htbp]
\centering
\includegraphics[width=\textwidth]{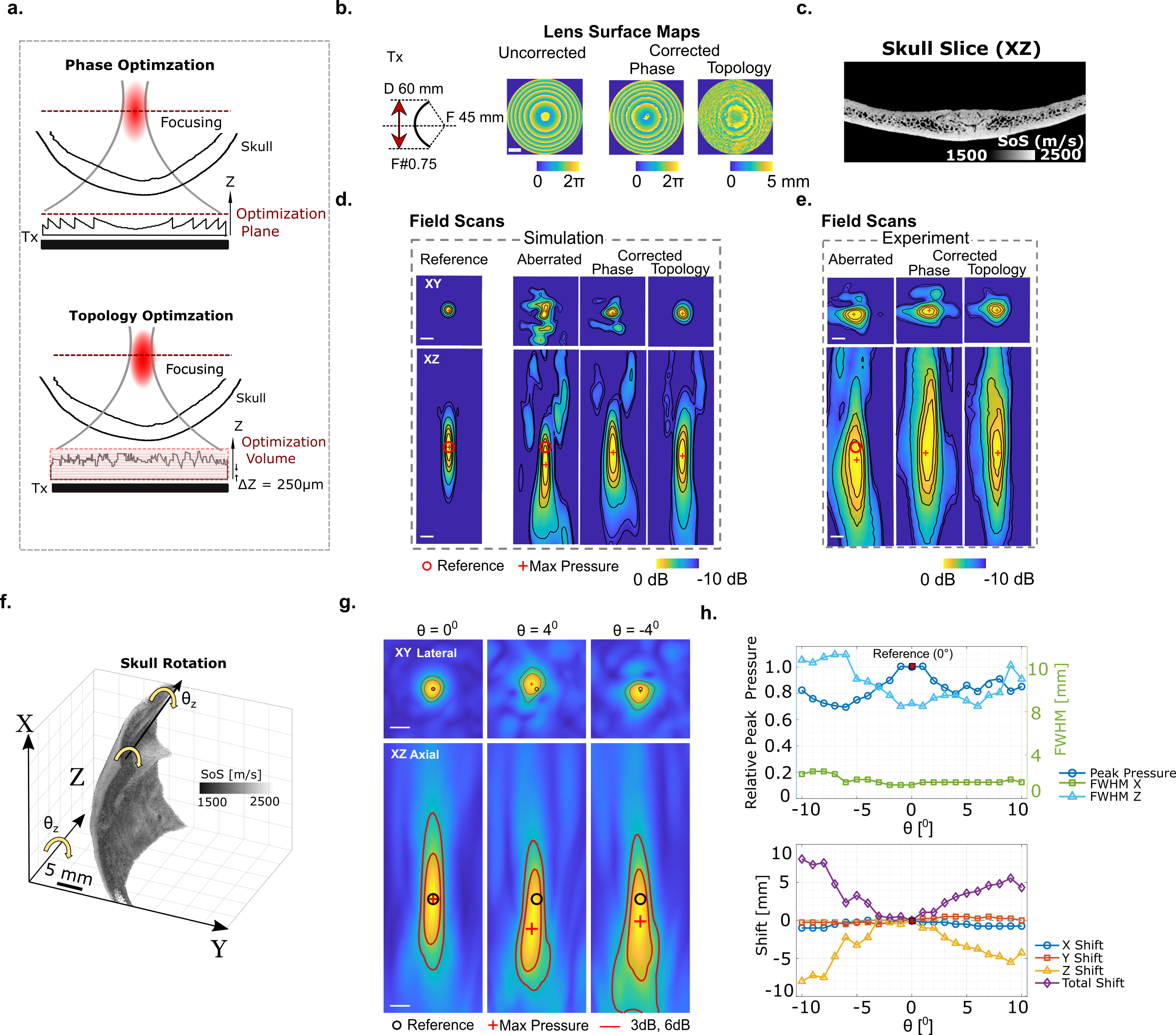}
\caption{Experimental validation of trans-skull hologram focusing and assessment of registration errors on focusing quality. (a) Schematic of phase and topology optimization for single focusing. (b) Single-point focusing phase maps for a transducer with a diameter of 60 mm and an $F\#$ of 0.75, with and without skull corrections (scale bar: 1 cm) and equivalent topology map. (c) Skull speed of sound map obtained from micro-CT scan. (d) Lateral and axial focal profiles with contours obtained from simulations and (e) experiments, both with and without aberration correction for phase and topology optimized lenses (scale bar: 2 mm). (f) Simulation mask for skull rotation (scale bar: 5 mm). (g) Axial and lateral 2D surface maps demonstrating targeting and focusing errors due to a $\pm 4°$ skull rotation (scale bar: 2 mm). (h) Effect of skull rotation on peak amplitude and FWHM (Top) and focal shift and x, y and z directions (Bottom).}
\label{fig:trans_skull}
\end{figure}

To demonstrate that HASA-ADAM thickness optimization can be used for TUS we designed optimized topologies for single focus (1 MHz with F-number 0.75) through human skull (Fig.~\ref{fig:trans_skull}a-b). We compared the thickness optimized lens with a lens design to generate single focus in free field and a lens optimized to account for aberration but designed with standard phase to thickness conversion (Fig.~\ref{fig:trans_skull}a-c). The pressure generated was compared using both acoustic simulations and ex vivo trans-skull experiments (Fig.~\ref{fig:trans_skull}d). We observed that the HASA-ADAM based framework not only corrected for aberration but also substantially reduced the sidelobes in both axial ($xz$) and lateral ($xy$) focal planes, as indicated by both experimental and simulated data (Fig.~\ref{fig:trans_skull}d). Additionally, the optimized hologram that incorporates aberration correction achieves a 24.5\% reduction in lateral 3 dB beam width and a 20\% increase in peak pressure as compared to focal spot optimization without aberration correction (Fig.~\ref{fig:trans_skull}d). Evidently the focal pressure using the other two lenses was characterized by significant aberration (Fig.~\ref{fig:trans_skull}d) and high side-lobes, demonstrating suboptimal performance for high frequency TUS.

The above data demonstrates the potential of the proposed framework to effectively correct skull-induced aberrations and lead to diffraction-limited performance, however they also indicate that the focus attained with the experimental system is below theoretical limits (Fig.~\ref{fig:trans_skull}e). This discrepancy between simulation and experiment is most likely either due to registration errors or uncertainties in the skull and lens material properties or a combination of them. Past investigations by us have demonstrated that skull properties need to deviate by more than 20\% to lead to significant errors~\cite{schoen2022experimental} (see also Fig. S1), which is many cases is unrealistic. Likewise, the lens material used in our investigations (resin) has been well characterized allowing to model the lens material with reasonable accuracy~\cite{bakaric2021measurement}. Thus, we turned our attention to skull-lens registration.

To better understand the role of skull-lens registration, we conducted additional simulations where we rotated the skull about the z-axis in the transverse plane at increments of $\pm 1°$ (Fig.~\ref{fig:trans_skull}f) and compared the baseline axial and lateral fields (Fig.~\ref{fig:trans_skull}g). By inspection, we see an observable increase in 3 dB FWHM area in lateral and axial directions (Fig.~\ref{fig:trans_skull}g and h-top), which are consistent with the observed degradation in the focusing performance shown in Fig.~\ref{fig:trans_skull}e. There is also a noticeable degradation in the focusing quality with skull rotation along with a significant shift axial (z) direction (Fig.~\ref{fig:trans_skull}g and h-bottom). Together, these findings (Fig.~\ref{fig:trans_skull}) highlight the potential of the optimized hologram design to effectively correct skull-induced aberrations and emphasize that merging effective lens design with accurate skull-lens alignment is critical for attaining optimal focus with high frequency systems ($\geq 0.7$ MHz).

\subsection{Skull-compensating lens misalignment augments nonlinear wave propagation and parametric array signal}

\begin{figure}[htbp]
\centering
\includegraphics[width=\textwidth]{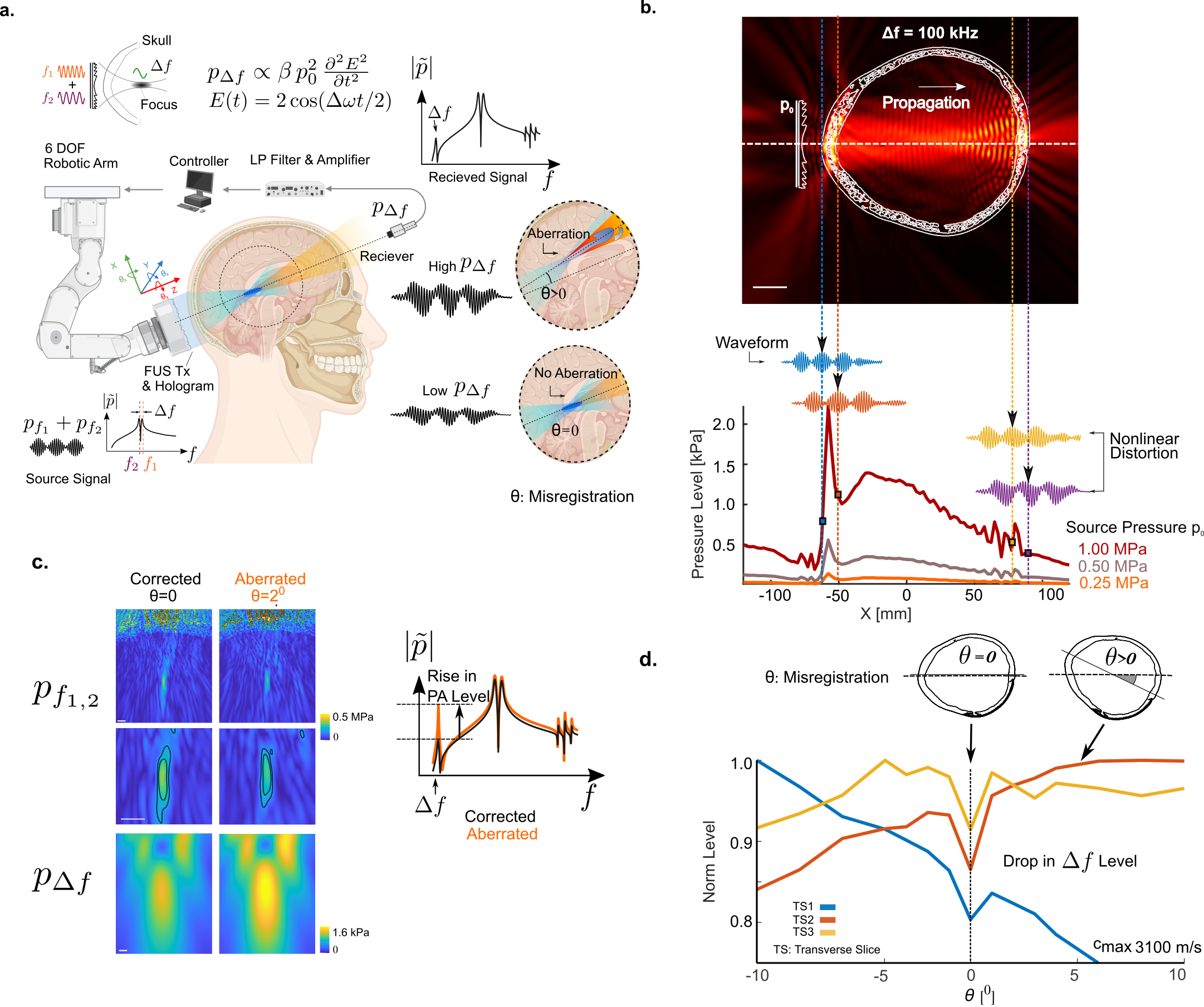}
\caption{Conceptual framework of skull-compensating lens registration using nonlinear acoustic feedback. (a) Top: Nonlinear mixing of two high-frequency waves generated by a flat ultrasound (US) transducer attached to a phase plate, resulting in focused ultrasound. At higher intensities, this process produces a low difference frequency. Bottom: Schematic illustration of the hologram-assisted FUS therapy device integrated with a six-degree-of-freedom (6 DOF) robotic arm. The system utilizes acoustic feedback based on nonlinear parametric array signals for accurate registration. (b) Top: 2D acoustic simulation demonstrating the generation of a 100kHz parametric field within the skull cavity and its subsequent transmission. Bottom: Waveforms showing increasing nonlinear distortion as primary waves attenuate due to skull-induced losses, and absolute pressure traces along the axial direction for varying source pressures. (c) Top: (Row 1), Simulation of the primary ultrasound field with 3dB (in red) and 6 dB (in black) contour maps in the skull region showing local pressure maxima in the skull for skull rotation $\theta=0°$ and $\theta=2°$, (Row 2) with zoomed view of the focal region as ROI with 3dB and 6dB contours (both in black); Bottom: The parametric field with ($\theta=2°$) and without ($\theta=0°$) skull rotation indicating aberration leads to higher PA signal (plot on the right). (d) 2D simulation results illustrate a decrease in parametric pressure corresponding to zero skull rotation ($\theta=0°$) for various transverse skull slices.}
\label{fig:nonlinear_concept}
\end{figure}

To address the long-standing challenge of skull-compensating lens registration and reconfigure this technology for high precision neuro-interventions, we turned our attention to largely unexplored non-linear wave propagation through human skull and the parametric array effect. Briefly, the parametric array effect is a nonlinear wave propagation effect~\cite{westervelt1963parametric}, where two (primary) high-frequency sound beams of finite amplitude interact to produce (secondary) sum and difference frequency beams. The strength of the difference frequency $|p_{\Delta f}|$, here termed parametric array signal (PA signal), which has several unique properties, including high directionality and penetration through the skull, is proportional to the medium's nonlinearity parameter $\beta$, the square of the primary beams amplitude $p_{f_{1,2}}$, and their propagation length~\cite{hamilton1997nonlinear}. Considering the characteristics of the PA signal, we hypothesized that skull-compensating lens mis-registration (i.e., suboptimal aberration correction) can affect the amplitude of $|p_{\Delta f}|$ that once detected and quantified (e.g., using a hydrophone) can provide a real-time feedback mechanism to noninvasively align the holographic lens to the patient's skull (Fig.~\ref{fig:nonlinear_concept}a).

To demonstrate that this effect can be observed in the brain, we employed nonlinear acoustic simulations using the k-wave toolbox~\cite{treeby2010k} along with bone and tissue nonlinearity parameters from the literature~\cite{renaud2008exploration,pinton2011effects}. Using primary frequencies of 0.95 MHz and 1.05 MHz that resulted in 100 kHz difference frequency and pressures ranging from 0.25 to 1 MPa (safe exposure), we found that the parametric signal was immediately evident after the primary beams passed through the skull (Fig.~\ref{fig:nonlinear_concept}b). Despite the formation of a weak standing wave ($\leq$1.5 kPa), which is expected due to the difference frequency used~\cite{baron2009simulation}, the parametric signal outside the skull was also evident and well within the detection limit of many piezoelectric detectors (Fig.~\ref{fig:nonlinear_concept}b, bottom). While the observed temporal profile is atypical of nonlinear propagation (i.e., the peak positive pressure tends to be higher), this is due to the accumulation of nonlinearities over extended propagation distances combined with the substantially higher attenuation of the primary and secondary MHz-range fields.

Building on these observations, we tested the impact of lens mis-registration on $|p_{\Delta f}|$. We found that in the presence of mis-registration the PA signal increases substantially (Fig.~\ref{fig:nonlinear_concept}c, plot on the right). Further analysis revealed that the aberrated beams, apart from distorting the pressure field (i.e., defocusing), also result in higher pressure buildup in the highly non-linear skull. This effectively extends the nonlinear interaction region, which is critical for the development of finite amplitude effects~\cite{hamilton1997nonlinear} (see also Eqn. 6 in Methods). Finally, we assessed the influence of misregistration on the PA signal by rotating the skull (Fig.~\ref{fig:nonlinear_concept}d). Interestingly, we found a steep increase in the PA signal for very small angles (i.e., small misregistration errors). Crucially, these observations persisted for different skull slices (Fig.~\ref{fig:nonlinear_concept}d), indicating that the PA signal drop is persistent and very sensitive to skull-compensating lens alignment. Together, these findings supported the notion that the low frequency acoustic signal generated by the nonlinear mixing of high frequency beams can be leveraged to attain accurate skull-compensating lens alignment.

\subsection{Sensitivity analysis reveals that parametric array signal is a robust and sensitive surrogate to skull-compensating lens alignment}

\begin{figure}[htbp]
\centering
\includegraphics[width=\textwidth]{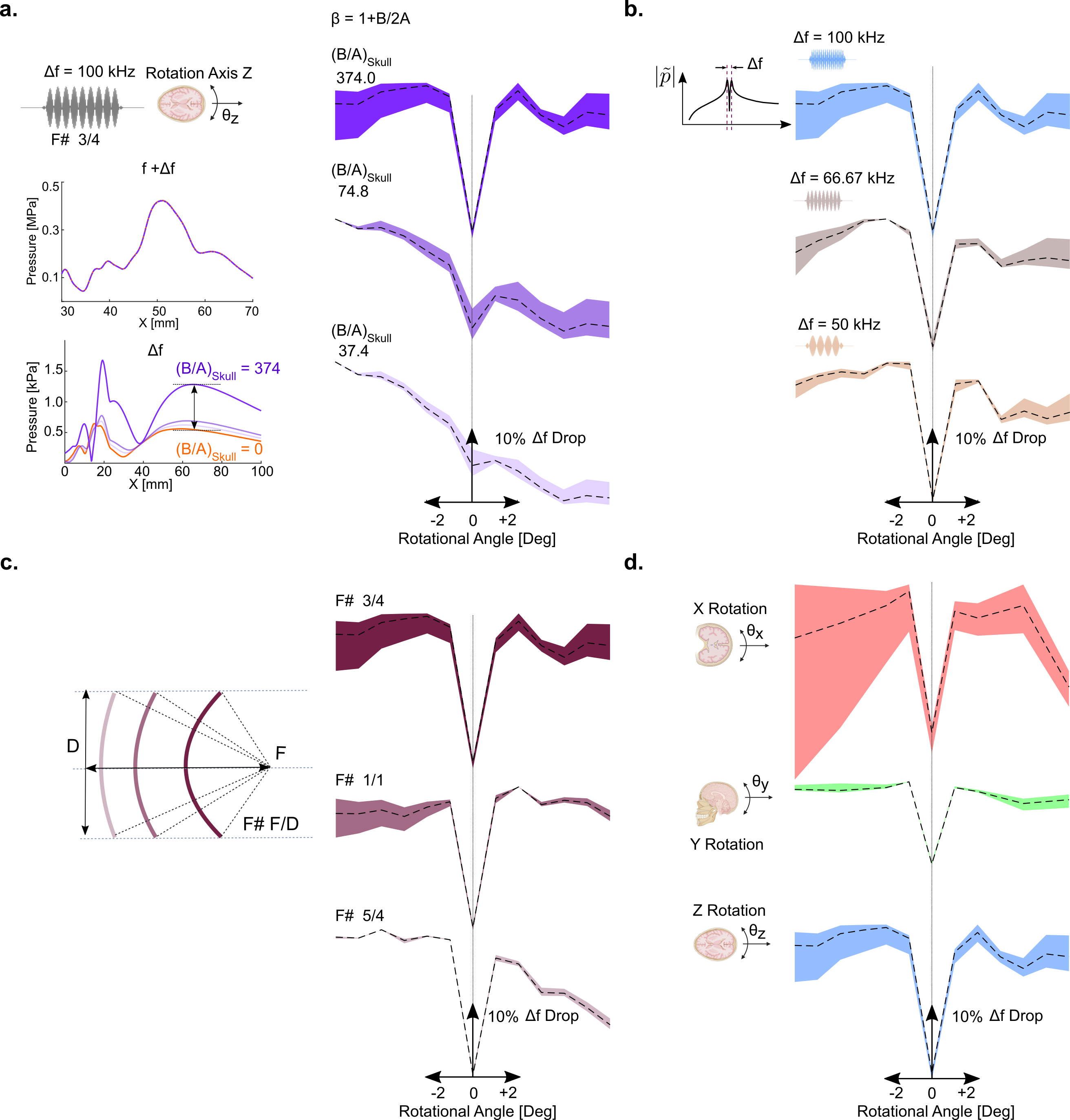}
\caption{Multiparametric sensitivity analysis reveals a robust and persistent drop in parametric pressure during optimal skull-compensating lens alignment. (a) Left: Primary and parametric array pressures under conditions of high and low skull nonlinearity. Right: Effect of varying skull nonlinearity on parametric pressure drop. The observed drop in parametric pressure indicates optimal registration of the hologram lens with the skull. (b) Influence of difference frequency on the parametric pressure. (c) Variation in parametric pressure drop with different depths of focusing (or F\#) across varying levels of nonlinearity, difference frequencies, and focusing parameters. (d) Impact of skull rotation about the x, y, and z axes on parametric pressure.}
\label{fig:sensitivity}
\end{figure}

Considering the above findings, we decided to delve deeper and conduct a more rigorous analysis using three-dimensional (3D) nonlinear simulations. First, we investigated the impact of skull nonlinearity $(B/A)_{\text{skull}}$ on parametric generation. As expected, the primary field (1.05 MHz) remained unaltered across different levels of skull nonlinearity, however the parametric field (100 kHz) decreased markedly when skull nonlinearity is absent (Fig.~\ref{fig:sensitivity}a). To further clarify this observation, we varied the skull nonlinearity parameter ($(B/A)_{\text{skull}} = 374, 74.8$, and 37.24; these are equivalent to Goldberg numbers of 3.0, 0.62, and 0.32, respectively; see methods for additional information) and performed multiple registration iterations by rotating the 3D skull in the transverse plane. Evidently, the parametric pressure drop followed closely the skull nonlinearity (Fig.~\ref{fig:sensitivity}a, right). Crucially, when skull misalignment is minimized (i.e. $\theta=0$) the drop in the PA signal becomes even more pronounced in the more realistic 3D simulations, as compared to 2D, for the same B/A parameters (i.e., Fig.~\ref{fig:sensitivity}a for $(B/A)_{\text{skull}} = 374$).

We also explored the effect of varying the difference frequency, $\Delta f$ (Fig.~\ref{fig:sensitivity}b) and observed that the PA signal drop appears to be insensitive to $\Delta f$, when it ranges from 50 -- 100 kHz. Notably, the relationship between PA signal amplitude and downshift ratio ($f/\Delta f$) follows established parametric array theory~\cite{westervelt1963parametric}, where larger downshift ratios yield smaller PA signals due to lower nonlinear interaction efficiency. Conversely, smaller downshift ratios, while potentially producing stronger signals, require transducers with broader bandwidth and are hindered by higher frequency-dependent attenuation through the propagation medium. This complex interplay of contributing factors determines the sensitivity of PA signal changes to downshift ratio variations. Next, we assessed the influence of focal depth by reducing the f-number while maintaining a constant aperture, producing a progressively weakly focused beam. Although we did not observe any major differences, lower f-numbers appear to have higher variation. Finally, the drop in the PA signal during optimal alignment is robust to different axis of rotation, although rotations about z-axis resulted in more substantial decreases in parametric pressure (Fig.~\ref{fig:sensitivity}d). Together, these findings demonstrate that the drop in the PA signal during optimal skull-compensating lens alignment is robust ($\geq$20\%) and persistent across a range of conditions, provided that the skull is characterized by high nonlinearity.

\subsection{Parametric array signal provides a real-time feedback mechanism to noninvasively align skull-compensating holographic lens to human skull}

\begin{figure}[htbp]
\centering
\includegraphics[width=0.8\textwidth]{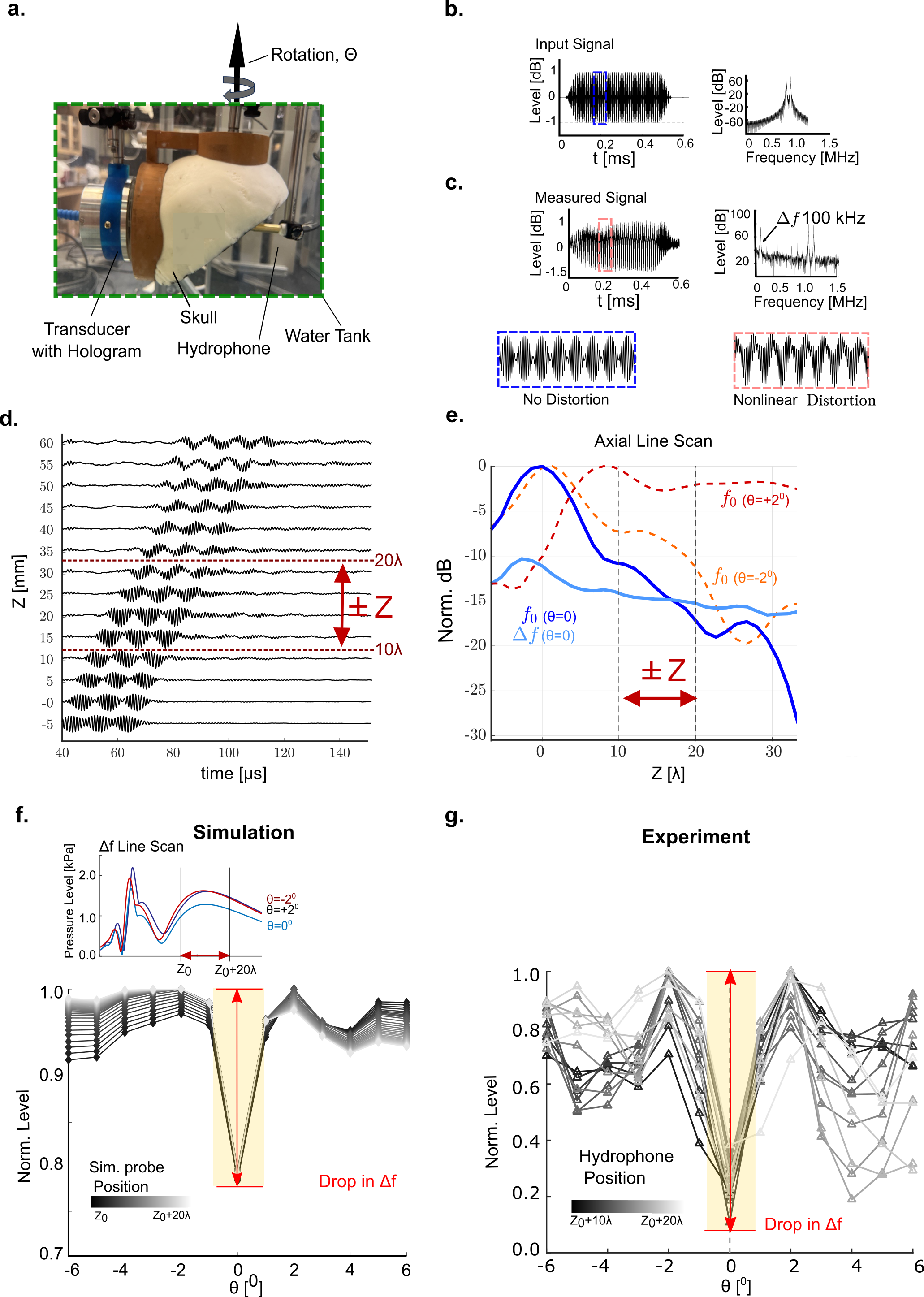}
\caption{Low frequency acoustic feedback generated by nonlinear mixing of high frequency waves enables accurate skull-compensating lens alignment ex vivo. (a) Experimental setup with the skull mounted on a rotating fixture. (b) Sample bi-frequency normalized input signal in both time and frequency domains; and Zoomed-in sections highlighting the absence of nonlinear distortion. (c) Sample measured signals using hydrophone and after 600kHz low pass filtering with 40 dB gain in both time and frequency domains; and Zoomed-in sections highlighting the presence of nonlinear distortion. (d) Stacked waterfall plot indicating progressive nonlinear distortion of the measured signal (for $\theta=0$). (e) Hydrophone line scans demonstrating Primary and parametric signals. (f) 3D simulation mimicking the experimental setup, illustrating the variation of parametric pressure with skull rotation in the transverse plane. (g) Experimental variation of parametric pressure, showing a decrease corresponding to zero registration error.}
\label{fig:experimental_validation}
\end{figure}

To validate the above findings, we designed a holographic lens using the HASA-ADAM framework (Fig.~\ref{fig:hasa_adam_framework}) and conducted experiments with 1 MHz transducer with active aperture D = 60 mm, coupled with a single focusing lens (F\# 0.75) the transducer was excited with bi-frequency input signal (containing 0.95 MHz and 1.05 MHz) at 0.2 MPa to produce 100kHz difference frequency. This signal was then recorded with a hydrophone and compared with three-dimensional simulations using the same geometry and skull segment (Fig.~\ref{fig:experimental_validation}). To be able to perform axial scans and characterize the PA signal at different distances from the skull using a hydrophone, we removed part of the skull (Fig.~\ref{fig:experimental_validation}a). Axial line scans confirmed that a 0.2 MPa (peak to peak) primary field (peak pressure at the focus) produced increasing nonlinear distortion along the axis and beyond the focal position. These are evident in the waterfall plot showing progressive self-demodulation (Fig.~\ref{fig:experimental_validation}d).

To minimize pseudo-sound effects that can appear in the measurements when the hydrophone is subject to strong primary pressure fields and bias (positively) our measurements (See also Suppl. Methods)~\cite{song2021experimental}, the measurement window was extended several wavelengths away from the focus (Fig.~\ref{fig:experimental_validation}d). After we established the presence of a measurable PA signal for clinically relevant primary pressures (M.I.= 0.1), we rotated the skull at $1°$ increments around the z-axis and measured its amplitude, as we did in the simulations above (Figs.~\ref{fig:nonlinear_concept} and \ref{fig:sensitivity}). The line scans for primary frequency revealed broadening of the axial beamwidth for both positive ($\theta>0$) and negative ($\theta<0$) registration errors (Fig.~\ref{fig:experimental_validation}e). These measurements, aggregated across z-axis positions, not only closely aligned with simulation predictions but also confirmed the pronounced drop in parametric pressure when the skull rotation was zero (Fig.~\ref{fig:experimental_validation}f-g). In aggregate, these findings (Figs.~\ref{fig:nonlinear_concept}-\ref{fig:experimental_validation}) support our hypothesis that the PA signal is very sensitive to skull-induced aberration and demonstrates its potential to provide a real-time feedback mechanism to align the skull-compensating holographic lens to the patient's skull.

\subsection{Holographic lens-based TUS and parametric pressure can monitor draining of excess cerebrospinal fluid away from the brain}

\begin{figure}[htbp]
\centering
\includegraphics[width=0.8\textwidth]{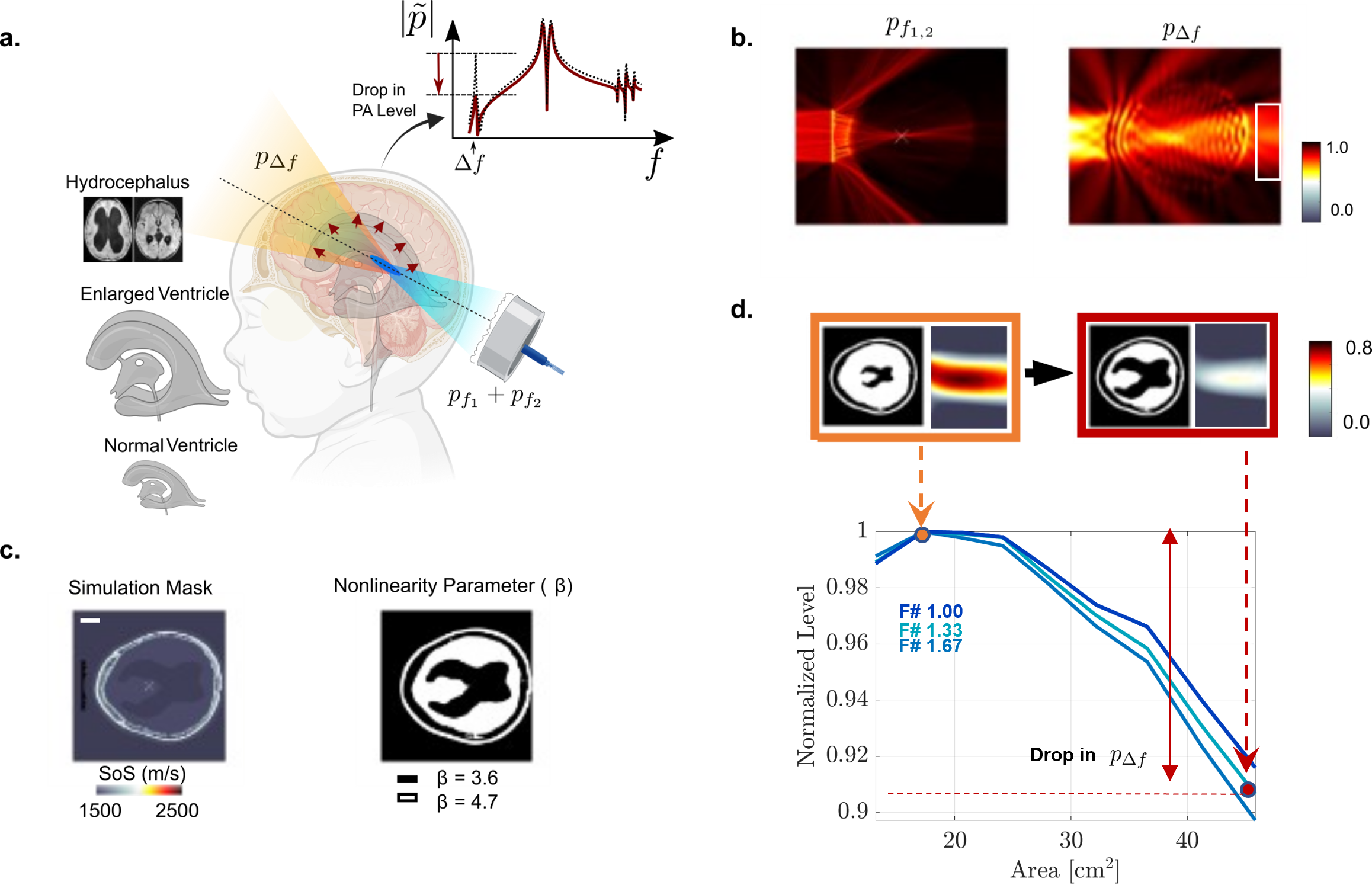}
\caption{Nonlinear acoustic feedback can detect relative changes in ventricular size. (a) Schematic showing hydrocephalus monitoring using parametric array effect, where increasing hydrocephalus leads to a drop in PA signal. (b) Simulation mask and nonlinearity parameter map correspond to enlarged hydrocephalus. (c) Primary and parametric field obtained from simulation. (d) Peak parametric signal pressure outside the skull cavity indicating a drop in parametric pressure with an increase in the size of ventricles.}
\label{fig:hydrocephalus}
\end{figure}

Inspired by these findings, we explored if nonlinear acoustic feedback can be also employed to monitor changes in the effective nonlinearity of the brain in response to relative changes in ventricular size during hydrocephalus progression or shunt treatment. Hydrocephalus is a condition that requires constant monitoring of intracranial pressure (ICP) as it is a key biomarker for monitoring disease progression and subsequent, removal of excess cerebrospinal fluid (CSF) (e.g. using shunt-based treatment). Although ICP monitoring using ventricular catheters is the gold standard, this approach is invasive and, as such, carries significant risks, including infection, hemorrhage, and neurological deficits~\cite{jiang2022invention}. Therefore, creating non-invasive techniques to confirm ICP change as well as monitor shunt function can reduce the risk of complications in hydrocephalus patients~\cite{fischer2020non}. Considering, that extended clinical investigations have demonstrated that removal of excess CSF leads to a significant drop in ICP that in turn results in reduced ventricular space (or increased space occupied by the brain tissue)~\cite{czosnyka2004monitoring}, we reasoned we can leverage the differences in nonlinear properties of brain tissue ($\beta = 6.6$) and CSF (assuming nonlinearity of water, $\beta = 5.2$) to infer changes in excess CSF~\cite{duck2013physical} (Fig.~\ref{fig:hydrocephalus}a).

To investigate the relationship between ventricular expansion and PA signal, we conducted 2D simulation studies, using the above methods and models (Fig.~\ref{fig:hydrocephalus}b). These simulations are designed to examine how clinically relevant variations in ventricular size influence the PA signal (Fig.~\ref{fig:hydrocephalus}c). Our simulations indicated that as the ventricular space doubles in size (progressively expands from 20 cm$^2$ to 40 cm$^2$ ventricular cross-sectional area), which is a clinically relevant scenario for untreated hydrocephalus~\cite{bendella2024brain}, the PA signal measured outside the skull cavity drops by 10\% (Fig.~\ref{fig:hydrocephalus}d). This increase is sensitive to the transducer F-number with tightly focused beams ($F\#1.00$) lead to a larger relative drop in parametric pressure. These findings demonstrate that nonlinear acoustic feedback can be used to noninvasively monitor changes in the effective nonlinearity of the brain along the primary propagation path in response to relative changes in ventricular size during hydrocephalus progression or ventricular contraction during shunt-based treatment.

\section{Discussion}

Our findings indicate that the HASA-ADAM framework generate holographic lens topologies that account for thickness-dependent factors such as amplitude errors, scattering within thicker holographic lenses and edge diffraction effects. This capability constitutes major advancement in acoustic holography as it allows us to generate holographic fields with fidelity that lenses based on thin-element approximation or realizable phased arrays cannot match (Fig.~\ref{fig:hasa_adam_framework}). Moreover, the automated differentiation through the ADAM optimizer reduces the risk of local minimum and maintains robust focal accuracy (Fig.~\ref{fig:trans_skull}). In terms of computational speed, HASA-ADAM bridges the gap between computationally intensive methods like Time-Reversal~\cite{sallam2024gradient,angla2023transcranial} and trained deep learning frameworks~\cite{bu2024deep,li2022acoustic,lee2022deep} that are nevertheless opaque and reliant on extensive training datasets. Although directly optimizing for thickness is more challenging compared to phase optimization and thus requires about twice as long to converge ($\sim$15 minutes at $\lambda/6$ discretization for 500 iterations), with improved tuning of the hyperparameter of optimization and initial conditions the convergence can be further accelerated. Likewise, hybrid strategies that merge the speed of deep learning methods with HASA's accuracy and interpretability of wave propagation with GPU acceleration can further mitigate the computational cost associated with the $O(N^2\log N)$ scaling allowing real-time implementations, which can be desirable in some applications~\cite{jiang2022flexible,naor2012towards}.

Despite the remarkable performance of the proposed framework to design holographic lens topologies, we observed some discrepancies among the optimization results, the k-wave simulations, and the hydrophone scans, especially in the high-fidelity holograms in free-field (Fig.~\ref{fig:hasa_adam_framework}c and d). These may originate from the following sources of error. First, the reported 2590 m/s group velocity~\cite{bakaric2021measurement} may not accurately represent the 3D-printed sample (using Clear white v4 resin) due to curing-induced density variations and internal stresses. Second, the manufacturing tolerances of $\pm$0.05-0.1 mm ($\sim$5\% of thickness)~\cite{lagerburg2025dimensional} may introduce additional timing errors. Another potential source of error is related to the speed of sound variations (e.g., from the lens to water), which for the current HASA implementation cannot be very high relative to wavelength~\cite{schoen2020heterogeneous}. These sources of error can be readily addressed with improved lens materials and material characterization. Finally, understanding the impact of multiple reflections, which are neglected in the currently implemented HASA algorithm, and their role on optimization convergence~\cite{stanziola2023physics}, may allow to further improve the hologram quality. Despite these potential sources of error, the proposed HASA-ADAM framework delivers a balanced approach for scalable, rapid, and accurate hologram-design that can produce high-fidelity holograms and accommodate patient-specific skull variations to support rigorous treatment planning for transcranial targeting of specific brain regions.

To comply with the targeting requirements in the brain, where mistargeting can have safety risks or when accurately targeting specific brain regions or neuronal circuits is of the essence~\cite{meng2021applications,airan2017neuromodulation}, a millimeter (i.e., sub-wavelength) targeting accuracy is needed. To achieve this level of accuracy our investigations addressed the long-standing challenge of skull-compensating lens registration, by uncovering the close relationship between skull nonlinearity and the aberrations caused by misregistration. This observation allowed to leverage low frequency acoustic feedback generated by the nonlinear mixing of high frequency beams as they propagate through the highly non-linear skull to precisely align with the patient's skull. This is beyond what current neuronavigational approaches can achieve, which is at best 2 mm~\cite{choi2024neuronavigation,chen2021neuronavigation,wei2013neuronavigation,pouliopoulos2020clinical}. Beyond its immediate application to skull-compensating lens registration, we envision that the PA acoustic feedback combined with the HASA-ADAM framework could also be utilized for noninvasive aberration correction of phased arrays, where it can be used as an objective function for noninvasive in vivo phase and amplitude optimization of each element. Together, these conceptual contributions and advancements support the design of simple, economical, and high-performance ultrasound systems for high precision neuro-interventions. Such systems may also support daily/weekly treatments, possibly at outpatient and/or limited resource settings, without compromising performance, thereby supporting the effective translation and broad dissemination (i.e., similar to US imaging) of this technology~\cite{schoen2022towards,rincon2022biomarkers}. It may also may alleviate the need for repeated use of intraoperative MRI during TUS interventions, such as targeted drug delivery or liquid biopsy, that can complicate or even prevent their implementation (e.g., the average time to get an MRI appointment can be several months~\cite{hofmann2023variations}).

While our experimental and numerical results indicate a substantial drop in the PA signal during good skull-compensating lens alignment, we noticed some discrepancies among them that can be attributed to several interrelated factors. First, the nonlinearity parameter $\beta$, which fundamentally governs PA signal generation, exhibits frequency-dependent behavior that is often oversimplified in simulations~\cite{panfilova2021review,zhang2001experimental}. Additionally, $\beta$ for skull, which is a highly porous structure, has not been characterized in literature, suggesting that current values used in theoretical investigations might not be optimal. Experimental uncertainties might also contribute. Most notably, microscopic air bubbles trapped in the skull pores may persist~\cite{tang2011effect}, despite the extended degassing we performed (see methods). These microbubbles that are resonant in the MHz range and exhibit extreme nonlinearity even at very low void fractions may contribute to skull non-linearity~\cite{cavaro2011microbubble,overvelde2010nonlinear}. Additionally, skull bone follows complex frequency-dependent attenuation and has high inter-individual variability that is often underestimated in simulations~\cite{pinton2012attenuation,pinton2011effects}. Together these sources of uncertainty can lead to higher Goldberg number (i.e., nonlinearities) and PA signal under experimental conditions. Considering the implications of our work, we anticipate that our findings will motivate research aimed at providing more detailed and accurate characterization of the linear and nonlinear properties of the skull.

Our study also illustrated the potential of monitoring relative changes in the effective nonlinearity parameter ($\beta$) of the brain/CSF in response to ventricular shrinkage (i.e., return to normal dimensions) associated with hydrocephalus shunt treatment. While, invasive ICP monitoring remains the gold standard~\cite{zhang2017invasive}, our findings complement a shift towards safer and noninvasive diagnostics that are critical for reducing infection and hemorrhage risks~\cite{geraldini2022transcranial,caricato2014echography}. By using inherent tissue properties it also offers better safety profile from contrast-enhanced ultrasound~\cite{zhang2021wearable}. In addition to experimental validation in large animal models (i.e., non-human primates), a more detailed understanding of the CSF/brain nonlinearity parameters ($\beta$) in health and disease along with beam geometry optimization will help advance this technology. Beyond its immediate implications for monitoring hydrocephalus shunt treatment, this approach, which leverages intrinsic tissue properties rather than external contrast agents, could also be used for early detection and disease progression, especially at resource-limited settings. Alternatively, it could be integrated to existing ultrasound protocols for diagnosing hydrocephalus in infants to improve their accuracy by mitigating challenges related to operator skill and need for acoustic windows~\cite{jiang2022invention,moskowitz2010cumulative,filippou2018recent}.

Collectively, the proposed research by accelerating hologram design and introducing robust registration strategies to support the design of high-fidelity transcranial holography, is poised to support the effective translation and broad dissemination of this promising technology to the clinics. While our work is primarily focused on biomedical applications, the implications of high-fidelity acoustic holography are much broader and will invite researchers to explore this new capability across a range of applications~\cite{melde2023compact,hirayama2019volumetric,xie2016acoustic,kruizinga2017compressive,jimenez2018adaptive,maimbourg2018printed,jimenez2019holograms,melde2016holograms}. Our research also lays the groundwork for future research exploring the low frequency nonlinear acoustic feedback for the diagnosis, monitoring, and treatment of brain diseases and highlights the importance of relatively thin and highly nonlinear media to augment finite-amplitude effects.

\section{Methods}

\subsection{Holographic Lens Design}

\subsubsection{Heterogeneous Wave Propagation with ASA}
The first step in our design process is to model wave propagation through skull heterogeneity. For a time-harmonic pressure field $\tilde{\bm{p}}(\bm{r})e^{-i\omega t}$, where $\omega$ is the angular frequency is, the angular spectrum $P$ is given by its 2D spatial Fourier transform

\begin{equation}
P(k_x, k_y, z) = \mathcal{F}_k[\tilde{p}(x, y, z)],
\end{equation}

For heterogeneous media where the spatial variation of sound speed $c(\bm{r})$ is less compared to the wavelength, the ordinary differential equation for the angular spectrum $P$ becomes

\begin{equation}
P_z + k_z^2P = \Lambda * P,
\end{equation}

Here, $\Lambda = \mathcal{F}_k[k_0^2(1 - \mu)]$, $k_0 = \omega/c_0$, $\mu = c_0^2/c^2$, $c_0$ is a reference (average) sound speed, and $*$ indicates two-dimensional convolution over the component wavenumbers $k_x$ and $k_y$. For our design, the skull and tissue densities were obtained from micro-CT scan data of a human skull~\cite{aubry2003experimental,arvanitis2015transcranial} (original resolution 95$\mu$m binned to 150$\mu$m which amounts to 10 points per wavelength for $f_0 = 1$ MHz and considering equilibrium sound speed $c_0 = 1480$ m/s in water). This wave propagation model captures refraction and transmission through the skull. In the current implementation absorption was not considered.

An implicit solution of Eq. 2 may be obtained with a Green's function technique~\cite{morse1953methods}, and numerical approximation allows computation of $P$ at arbitrary $z$ via

\begin{equation}
P^{n+1} \approx P^n e^{ik_z\Delta z} + \frac{e^{ik_z\Delta z}}{2ik_z}(P^n * \Lambda) \times \Delta z,
\end{equation}

where $P^n = P(k_x, k_y, n\Delta z)$.

Provided the marching step size $\Delta z$ is much smaller than the wavelength ($\lambda$), Eq. 3 enables calculation of the field in the heterogeneous medium. We have chosen $\Delta z = 250\mu$m here which is $\lambda/6$. Using the above wave propagator, pressure distribution at the target plane $P_t$ can be efficiently computed from initial pressure field $P_0$.

\subsubsection{Approach 1 - Topology Optimization}
Topology optimization directly optimizes the material thickness distribution to achieve a desired intensity profile at the target plane. This method employs the HASA Algorithm to model skull heterogeneity and acoustic propagation through complex media. The algorithm initializes with a uniform thickness distribution $\bm{u}^{(0)}$ and iteratively refines it to minimize the loss function. At each iteration $n$, the thickness is computed as:

\begin{equation}
t^{(n)} = \min\{\text{softplus}(u^{(n)}), t_{\max}\},
\end{equation}

where the softplus function ensures non-negative thickness values and $t_{\max} = 5$ mm imposes an upper bound on the hologram thickness. The complex pressure field $P^{(n)}$ is then calculated using the HASA forward model $\mathcal{H}(P_0, t^{(n)})$, where $P_0$ represents the transducer's initial pressure distribution. The loss function quantifies the mismatch between the computed intensity $I^{(n)} = |P^{(n)}|^2$ and the target intensity $I_t$ at each spatial location $(x,y)$:

\begin{equation}
\mathcal{L}^{(n)} = \sum_{x,y} |I^{(n)}(x, y) - I_t(x, y)| + \lambda\mathcal{R}(t^{(n)}),
\end{equation}

where $\lambda\mathcal{R}(t^{(n)})$ is a regularization term that encourages smooth thickness variations, with $\lambda = 0.01$ controlling the regularization strength. The optimization employs the ADAM optimizer with learning rate $\eta = 0.01$ to update the thickness parameter:

\begin{equation}
u^{(n+1)} = u^{(n)} - \eta\nabla_u\mathcal{L}^{(n)},
\end{equation}

The gradients are computed using automatic differentiation within TensorFlow. The iterative process continues until either the loss converges below a threshold $\epsilon = 0.001$ or the maximum number of iterations $N_{\max}$ is reached (500 for a single point focusing, 2000 for complex 2D distributions). The hologram thickness optimization steps are summarized in Table~\ref{tab:algorithm}.

\subsubsection{Approach 2 - Phase Optimization}
The phase optimization approach iteratively refines the phase distribution $\phi$ at the hologram surface while maintaining a fixed amplitude profile. The algorithm begins with an initial uniform phase distribution $\phi^{(0)}$ and zero thickness assumption. At each iteration, the algorithm computes the pressure field using:

\begin{equation}
P_t^{(n)} = \mathcal{H}(P_0, \phi^{(n)}),
\end{equation}

where $\mathcal{H}$ represents the HASA propagation operator. The intensity at the target plane is calculated as $I_t^{(n)} = |P_t^{(n)}|^2$. The loss function for phase optimization takes the form:

\begin{equation}
\mathcal{L}^{(n)} = \sum_{x,y} |I_t^{(n)} - I_t| + \lambda\mathcal{R}(\phi^{(n)}),
\end{equation}

where $I_t$ is the target intensity distribution and $\mathcal{R}(\phi^{(n)})$ regularizes the phase profile to encourage smoothness. The phase is then updated using gradient descent with the ADAM optimizer. Following convergence, the optimized phase profile is converted to a thickness map. Furthermore, the optimization routine considers the impact of hologram thickness on the amplitude transmission. This is achieved by converting the refined phase profile at the transducer surface into a thickness map. This conversion relies on the relationship between phase change and thickness variation, represented by:

\begin{equation}
\Delta\phi(x,y) = (k_w - k_h)\Delta h(x,y),
\end{equation}

Here, $k_w$ and $k_h$ denote the wave numbers for water and the hologram material, respectively. The thickness in turn used to compute the transmission coefficient $\alpha_T$ and the complex amplitude at the hologram plane following established expressions~\cite{melde2016holograms}. The hologram phase optimization steps are summarized in Table~\ref{tab:algorithm}. Effective implementation and parameter tuning are crucial for optimal results.

\subsubsection{Fabrication}
The last step in the design process is the fabrication of holograms. For topology optimization (i.e. approach 1) we 3D-print the thickness mask without any additional processing. To do this for phase optimization (i.e. approach 2) we utilize equation (9) to convert the optimized phase map to a thickness map and 3D-print it. In our design, we have used clear white v4 resin from Formlabs (Somerville, MA). This resin has low attenuation values across the frequency range of interest and higher greater speed of sound (with group velocity $c_g = 2591$ m/s and attenuation, $\alpha_0 = 2.922$ dBMHz$^{-1.044}$ cm$^{-1}$) making it suitable for 3D printing acoustic holograms among the materials characterized by Bakaric et. al.~\cite{bakaric2021measurement}.

\subsection{Hologram Design Validation}

\subsubsection{Trans-Skull Simulations}
Three-dimensional linear acoustics simulations were performed using open-source ultrasound simulation toolbox k-wave~\cite{treeby2010k} with GPU acceleration. Elastic effects were ignored as the angle of incidence on the skull bone in our study was below the critical angle in most cases. All the simulations were performed with a spatial grid size of $\Delta x = \Delta y = 250\mu$m and a CFL number less than 0.3 to ensure numerical stability. We used a porosity-dependent semi-empirical relationship to convert Hounsfield unit $H$ obtained from micro-CT scan into medium properties (such as speed of sound $c$, density $\rho$, and attenuation $\alpha$). The maximum speed and density of solid bone were taken as $c_{\text{bone}} = 2500$ m/s and $\rho_{\text{bone}} = 2000$ kg/m$^3$ respectively. Spatially dependent attenuation $\alpha$ was assigned using a power law absorption $\alpha = \alpha_0 \cdot f^{\beta}$ with uniform exponent $\beta = 1.2$.

\subsubsection{Trans-Skull Experiments}
The ex-vivo trans-skull experiments were performed in a degassed deionized water tank setup. A 60 mm diameter piston transducer (Precision Acoustics, Dorchester, UK) was coupled to 3D printed hologram lens. The transducer and lens assembly were then attached to the parietal region of an overnight degassed (approximately 12h) skull cap (Skulls Unlimited, Oklahoma City, OK, USA). The focal field of the lens was scanned with a calibrated 2 mm needle hydrophone (Precision Acoustics, Dorchester, UK) mounted on a three-axis positioning system (Velmex, Bloomfield, NY, USA) and was recorded by a digital oscilloscope (Pico Technologies, St Neots, UK).

\subsection{Hologram Registration Approach}

\subsubsection{Leveraging Nonlinear Parametric Array Effect}
The parametric array effect is a phenomenon in which nonlinear interaction between two high frequency sound beams $f_1$ and $f_2$ generates sum and difference frequencies $f_{\pm} = f_1 \pm f_2$ under finite-amplitude (i.e., high-intensity) wave propagation. Here, the parametric difference frequency $f_-$ is written as $\Delta f$. This occurs when collimated sound beams with slightly different frequencies interact nonlinearly~\cite{ingard1955scattering,westervelt1957scattering}, the difference frequency generation-termed the parametric array effect-enables focusing of low frequency pressure with sub-wavelength precision, which has been explored for the inference of tissue stiffness in elastography~\cite{fatemi1998ultrasound,karjalainen1999ultrasound,konofagou2001focused}. The parametric pressure at the difference frequency can be expressed via the modulating envelope $E(t) = 2\cos(\Delta\omega t/2)$~\cite{pompei2002sound,berktay1965possible}, as

\begin{equation}
p_{\Delta f} \propto \beta p_0^2 \frac{d^2E^2}{dt^2},
\end{equation}

where $p_0$ and $p_{\Delta f}$ are the primary and difference frequency waves, and $\beta$ is the nonlinearity parameter and $\omega = 2\pi f$ is the angular frequency. This relation holds for both planar and focused sources, with the difference frequency amplitude directly proportional to the primary amplitude and medium nonlinearity. We utilize the proportionality $p_{\Delta f} \propto \beta p_0^2$ to detect lens misregistration relative to the skull. By driving the transducer with two primary waves around a high frequency (e.g., 1 MHz) differing by a low difference frequency (e.g., 100 kHz), we focus the difference frequency in the brain within the weakly nonlinear regime. The nonlinearity parameter $\beta$ for high-density trabecular bone using B/A = 374~\cite{renaud2008exploration} and $\beta = 1 + \frac{B}{2A}$, ($\beta_{\text{bone}} = 188$) is much higher than that for coupling media like water ($\beta_{\text{water}} = 3.5$) and brain ($\beta_{\text{brain}} = 4.45$). Thus, variations in the parametric pressure is sensitive to skull-lens misregistration.

\subsection{Hologram Registration Validation}

We employed nonlinear ultrasound simulations to simulate the impact of skull bone misregistration on the parametric field. Specifically, by solving second-order wave equation known as Westervelt's equation, which accounts for nonlinear propagation with thermoviscous effects. For this purpose, we utilized a pseudo-spectral time domain (PSTD) approximation of Westervelt's equation implemented in the k-wave toolbox. Although the inclusion of higher-order harmonics due to nonlinearity significantly increases computational overhead, this challenge was effectively addressed through GPU-enabled parallelization. Power law absorption was modeled using the fractional Laplacian operator $(\nabla^2)^{y/2}$~\cite{chen2004fractional}. Additionally, nonlinearity was incorporated through the following lossless discrete pressure-density relation~\cite{treeby2010k}:

\begin{equation}
p^{n+1} = c_0^2\left[\rho^{n+1} + \frac{B}{2A}\frac{1}{\rho_0}(\rho^{n+1})^2\right],
\end{equation}

where $\rho^n = \rho(\bm{r}, n\Delta t)$. The nonlinearity parameter $\beta$ for water and skull bone were kept at 3.5 and 374 respectively. The objective was to measure the difference frequency field near the focal region under conditions of misregistration between the skull bone and the lens. Skull misregistration was induced by rotating the skull in the transverse plane (i.e., about the $z$-axis) by $\pm 1°$, and the difference frequency field was subsequently measured at 100 kHz. In the literature, the nonlinearity parameter $\beta$ for bone, which is not well-characterized, is generally ignored, due high attenuation and thin cross-section of the skull bone~\cite{rosnitskiy2019simulation}. To gain further insights, simulations were conducted by varying the nonlinearity parameter $\beta$ and the absorption coefficient $\alpha$. These parameters were combined using Goldberg's number $\Gamma$, defined as $\Gamma = \frac{\epsilon\beta}{\alpha/k}$ where $k$ is the wave number and $\epsilon$ is the acoustic Mach number, defined by $\epsilon = \frac{p_0}{\rho_0c_0^2}$. Essentially, the Goldberg number represents the ratio of the strength of nonlinear effects to the attenuation over one wavelength.
\newpage
\section*{Acknowledgments}

\textbf{General:} We thank Dr. Stas Emelianov for providing the E\&I high-power amplifier. We also thank Drs. S. Schoen Jr. and L. Degertekin, as well as the LEAP HI team, for their valuable input, especially during the early phase of this research.\\
\textbf{Funding:} This study was supported by National Science Foundation (Division of Civil, Mechanical and Manufacturing Innovation, Leading Engineering for America's Prosperity, Health, and Infrastructure) Grants 1933158 and 1830577 (P.D., C.A.).\\
\textbf{Author contributions:} P.D. and C.A. designed research; P.D. and C.A. performed research; C.A. supervised the project; P.D. analyzed data; P.D. and C.A. wrote the paper.\\
\textbf{Competing interests:} The authors declare no competing interests.\\
\textbf{Data and materials availability:} All data needed to evaluate the conclusions of this study are present in the paper and the Supplementary Materials.


\bibliographystyle{ieeetr}  
\bibliography{refs}

\newpage
\section*{Supplemental Information}

\section*{Acoustic holography in the megahertz frequency with optimal lens topologies and nonlinear acoustic feedback}

\textbf{Pradosh P. Dash}$^{1*}$ \textbf{and Costas D. Arvanitis}$^{1,2}$\\
$^1$Woodruff School of Mechanical Engineering, Georgia Institute of Technology, Atlanta\\
$^2$Coulter Department of Biomedical Engineering, Georgia Institute of Technology and Emory University, Atlanta

\section{A. Sound-speed (SOS) and Frequency variation}

Errors in estimation of speed of sound of the skull can lead to uncertainty in aberration correction. Since precise knowledge of skull acoustic properties is challenging to obtain under clinical conditions, it is important to understand how these uncertainties propagate through holographic reconstruction along with their effect on focusing accuracy. While our main analysis assumes operation at the design frequency, real transducers have finite bandwidth and may operate at frequencies that deviate from the nominal design value, we also assessed the effect of deviation from the design frequency on focusing performance.

\textbf{Speed of Sound (SOS) variation:} A phase-only hologram is designed for a skull speed $c_{\text{design}} = 2500$ m/s and average thickness $d_{\text{skull}} = 7$ mm. If the actual speed is $c = c_{\text{design}}(1 \pm 0.15)$, the one-way travel-time error is

\begin{equation}
\Delta t = \left(\frac{1}{c_0} - \frac{1}{c}\right)d_{\text{skull}} \approx \frac{\pm 0.15 d_{\text{skull}}}{c_0} = \pm 4.2 \times 10^{-7}\text{s},
\end{equation}

At the design frequency $f_0 = 1$ MHz this corresponds to a phase slip

\begin{equation}
|\Delta\phi| = 2\pi f_0|\Delta t| \approx 2.64\text{rad}(151°),
\end{equation}

Using simulations, we varied the speed of sound of the skull for two focusing configurations with phase only lenses computed from time of flight for Focus at 45 mm ($F\#0.75$) and 60 mm ($F\#1.0$) and observed the effect of peak amplitude and peak location variation. We see about ($\pm 30\%$) variation in peak focal pressure while the axial drift ranges are relatively minor (Fig S1 Right). This is evident from the fact that SOS error ($\pm 15\%$) results in a max shift in focus (2.0 mm for x45 configuration, Table S1), which slightly exceeds the wavelength in water at 1 MHz (1.5 mm). For application that require high precision targeting this can still be a significant source of error and methods to account for it should be considered.

\begin{figure}[H]
\centering
\includegraphics[width=1.0\textwidth]{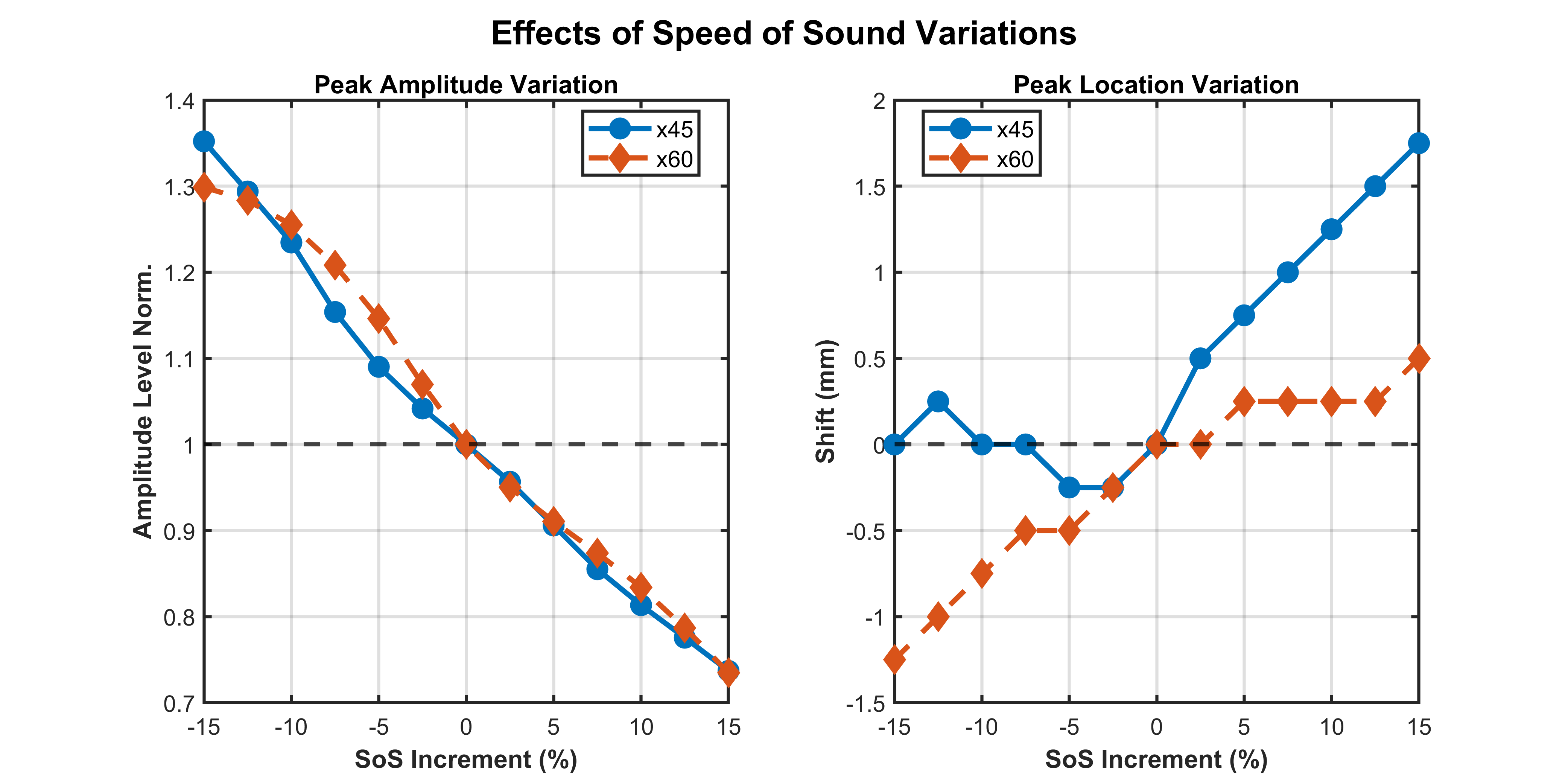}
\caption{Peak pressure and peak location versus SOS error ($\pm 15\%$) for both nominal foci.}
\label{fig:s1}
\end{figure}

\begin{table}[H]
\centering
\caption{Peak Amplitude and Location Statistics for Two Configurations}
\label{table:s1}
\begin{tabular}{lccccc}
\hline
Configuration & Min & Max & Range & Mean & Std Dev \\
\hline
\multicolumn{6}{c}{Peak Amplitude (Norm.)} \\
x45 & 0.737 & 1.352 & 0.615 & 1.016 & 0.200 \\
x60 & 0.735 & 1.299 & 0.564 & 1.027 & 0.197 \\
\multicolumn{6}{c}{Peak Location (mm)} \\
x45 & -0.250 & 1.750 & 2.000 & 0.500 & 0.685 \\
x60 & -1.250 & 0.500 & 1.750 & -0.212 & 0.548 \\
\hline
\end{tabular}
\end{table}

\textbf{Frequency variation:} To keep our analysis simple, we assume a homogenous medium. A planar aperture is driven with a static phase pattern

\begin{equation}
\phi_c(r) = -\frac{2\pi f_0}{c}\left(\sqrt{z_0^2 + r^2} - z_0\right),
\end{equation}

wrapped into the range $[0, 2\pi)$. Applying the same pattern at a new frequency, $f$ produces an effective time delay

\begin{equation}
\tau'(r) = \frac{\phi_c(r)}{2\pi f} = \frac{f_0}{f}\frac{\sqrt{z_0^2 + r^2} - z_0}{c},
\end{equation}

so the quadratic phase coefficient becomes $(f_0/f)$ times smaller. The new on-axis focus $z(f)$ satisfies

\begin{equation}
z(f) = \frac{f_0}{f}z_0,
\end{equation}

Using simulations, we varied the frequency of excitation for two focusing configurations with phase-only lenses computed from the time of flight for focus at 45 mm ($F\#0.75$) and 60 mm ($F\#1.0$) and observed the effect on focusing. The wavelength changes scale the beamwidth and depth-of-field inversely with $f$; high frequency tightens and attenuates the beam, whereas low frequency broadens and deepens it, moving the focus away from its intended position (Eqn. S5). The analysis suggests that for $\pm 5\%$ frequency shifts (which is the range of our registration trial with $f \pm \Delta f/2$ the max shift is comparable to 2-3 wavelengths in water at 1 MHz (Fig S2 bottom right, and Table S2). This is not necessarily a limitation of our registration strategy as the initial shift due to varying frequency is present in all the rotation configurations equally, and our method relies on relative changes between configurations rather than absolute positioning.

\begin{figure}[H]
\centering
\includegraphics[width=\textwidth]{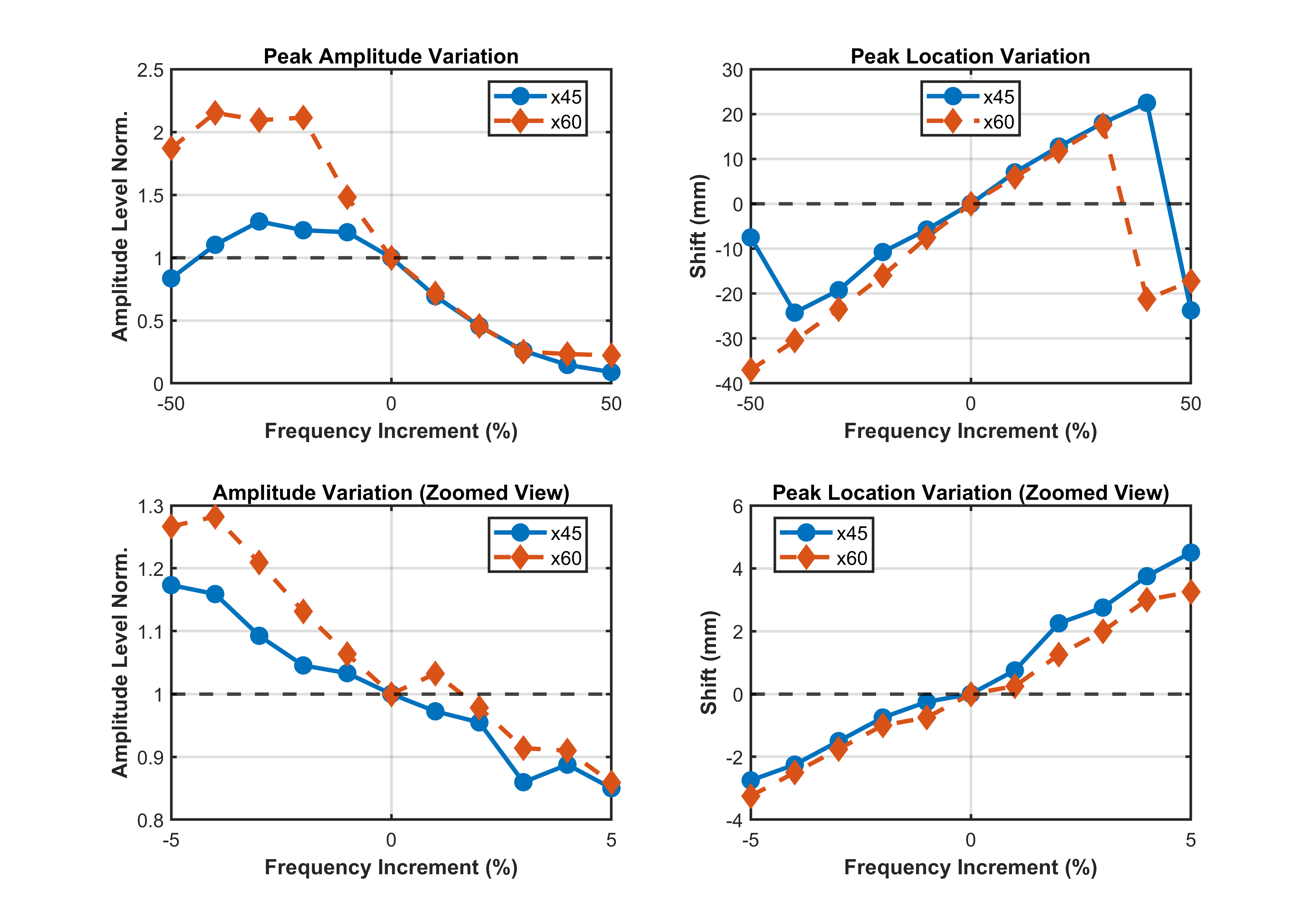}
\caption{Peak pressure and location versus frequency error ($\pm 50\%$) for both nominal foci.}
\label{fig:s2}
\end{figure}

\begin{table}[H]
\centering
\caption{Peak Amplitude and Location Statistics}
\label{table:s2}
\begin{tabular}{lccccc}
\hline
Configuration & Min & Max & Range & Std Dev & Std Dev \\
\hline
\multicolumn{6}{c}{Peak Amplitude (Norm.)} \\
x45 & 0.850 & 1.173 & 0.323 & 1.003 & 0.112 \\
x60 & 0.859 & 1.282 & 0.423 & 1.059 & 0.147 \\
\multicolumn{6}{c}{Peak Location (mm)} \\
x45 & -2.750 & 4.500 & 7.250 & 0.591 & 2.432 \\
x60 & -3.250 & 3.250 & 6.500 & 0.045 & 2.159 \\
\hline
\end{tabular}
\end{table}

\section{B. Pseudo-Sound in Parametric Arrays}

Pseudo-sound is a spurious low-frequency signal that appears in parametric arrays but does not originate in the acoustic medium itself. Instead, it arises from nonlinearities in the hydrophone and/or the receiving electronics. When two primary high-frequency signals, $f_1$ and $f_2$, are transmitted, the acoustic pressure can be modeled as:

\begin{equation}
p = P_0 \cos\left(\frac{\Delta\omega t}{2}\right)\cos(\omega_0 t),
\end{equation}

where $\omega_0$ is the carrier frequency, and $\Delta\omega = \omega_1 - \omega_2$ is the difference frequency. Although a genuine difference-frequency wave is generated in the medium through the nonlinear interaction of the primary waves, an additional low-frequency signal can appear due to the hydrophone's nonlinear transfer function.

\textbf{Nonlinear Transfer Function of the Hydrophone:} A common way to express the hydrophone's approximate transfer function is:

\begin{equation}
e \approx mp + \eta p^2,
\end{equation}

where $p$ is the incident pressure, $m$ is the linear sensitivity, and $\eta$ captures the hydrophone's nonlinear behavior. Since $p$ contains both the carrier and the difference-frequency components, the $p^2$ term can lead to unintended low-frequency outputs - often referred to as pseudo-sound. When expanded, the square-law term produces:

\begin{equation}
e \sim \eta p^2 \rightarrow \eta P_0^2 \times [\text{DC and combination frequencies}],
\end{equation}

giving rise to both a DC offset and low-frequency signals at or near $\Delta f$.

\textbf{Experimental Quantification:} Various strategies can quantify pseudo-sound: near-field measurements taken close to the transducer emphasize pseudo-sound because the true parametric wave has little distance to form; short-path suppression, achieved by minimizing propagation distance in water (e.g., a small tank), diminishes the genuine difference-frequency wave and helps isolate pseudo-sound; comparing signals over multiple distances exploits the fact that real difference-frequency waves scale differently with range than pseudo-sound; and using truncators to partially block or attenuate the primary beams lowers overall amplitude and thus square-law effects. We took advantage of the near-field measurements to quantify and isolate the pseudo-sound. Figure S3 shows the difference frequency at $\Delta f = 100$ kHz measured using a needle hydrophone as a function of primary pressure $\Delta f_1 = 1.05$ MHz for three separate conditions: at focus, at 60 mm from focus, and at 120 mm from focus. Evidently, as the hydrophone goes away from the focus, the primary pressure drops down (indicated by the rightward shift in the curves), however, the difference frequency pressure at 60 mm and 120 mm from focus remains the same and has almost linear relationship with the primary pressure. On the other hand, the difference frequency pressure at the focus has a quadratic relationship with the focal pressure (i.e. the quadratic term that is associated with pseudo-sound is significant). Since the PA signal remains constant while primary pressure decreases, this indicates the measured signal originates from pseudo-sound rather than true parametric generation. Consequently, to minimize pseudo-sound effects in our skull registration experiments, we performed all difference-frequency pressure measurements at locations away from the focal point, ensuring our results reflect genuine parametric phenomena rather than measurement artifacts.

\begin{figure}[H]
\centering
\includegraphics[width=1.0\textwidth]{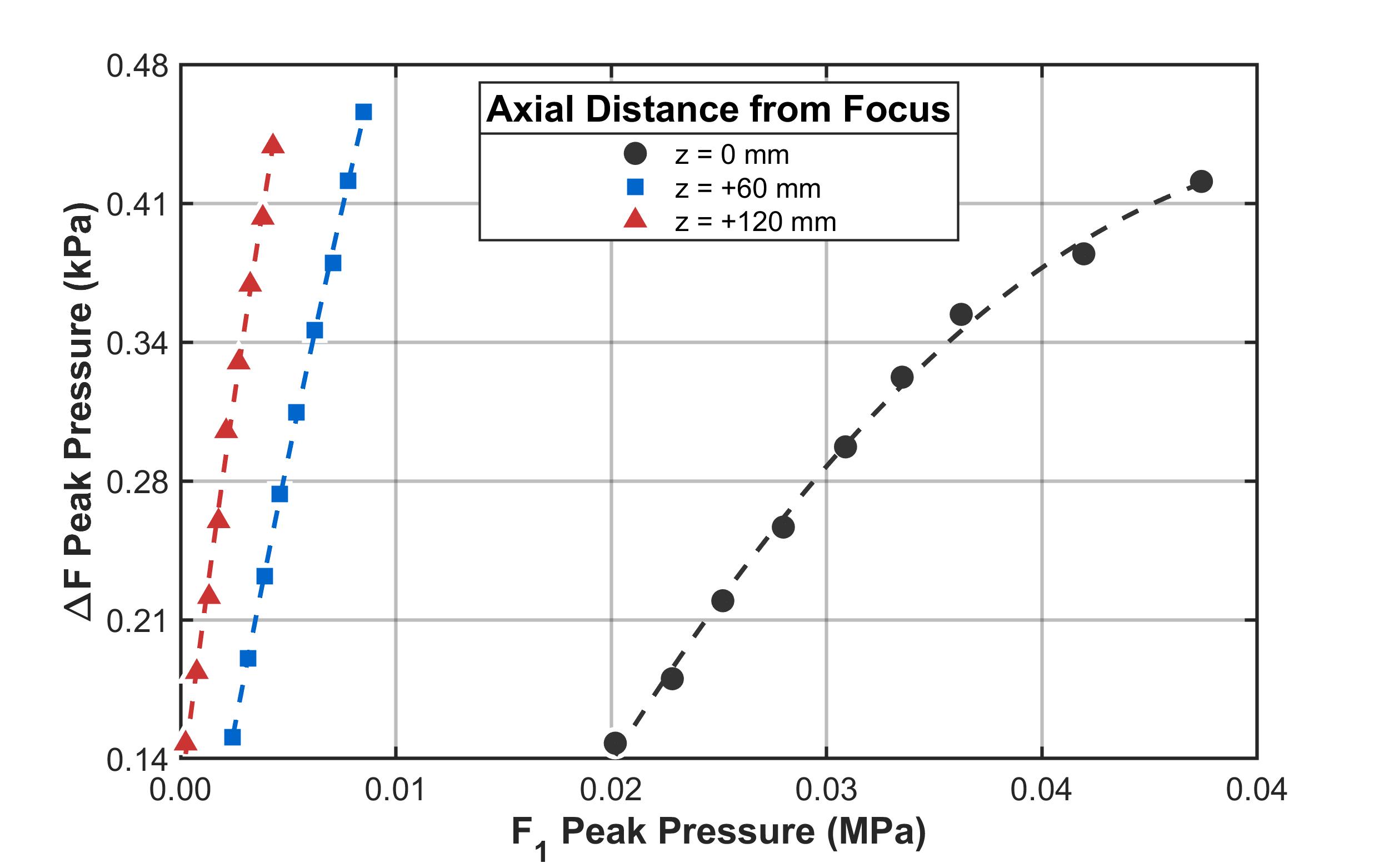}
\caption{Pseudo-sound quantification using near-field measurements. Axial Distance from Focus: z = 0 mm (black circles), z = +60 mm (blue squares), z = +120 mm (red triangles).}
\label{fig:s3}
\end{figure}

\section*{References}

\begin{enumerate}
\item J. Song, D. Jung, J. S. Kim, J. Lee, Experimental evaluation of pseudo-sound in a parametric array. \textit{J. Acoust. Soc. Am.} \textbf{150}, 3787–3796 (2021).

\item M. B. Moffett, J. E. Blue, Hydrophone nonlinearity measurements. \textit{J. Acoust. Soc. Am.} \textbf{68}, S35 (2005).

\end{enumerate}

\end{document}